%Paper: hep-th/9303107
%From: <MICUM%ROIFA.BITNET@pucc.Princeton.EDU>
%Date: Fri, 19 Mar 1993 10:32 +0200

The $SU_{p,k}(2)$ quantum algebra with two parameters $p$ and
$k$ [1] is defined by three generators $J_0,J_\pm$ with the
following commutation relations
$$J_0~J_\pm~-~J_\pm~J_0~=~\pm~J_\pm\eqno(1)$$
$$J_+~J_-~-~{p\over k}~J_-~J_+~=~[[~2~J_0~]]\eqno(2)$$
and the hermitian properties
$$J_0^+~=~J_0~,~(J_\pm)^+~=~J_\mp\eqno(3)$$
where
$$[[x]]~=~{k^x~-~p^{-x}\over k-p^{-1}}~.\eqno(4)$$

This algebra can be obtained from a single parameter $SU_q~(2)$
algebra defined by three generators $L_0~,~L_\pm$ having the
commutation relations
$$L_0~L_\pm~-~L_\pm~L_0~=~\pm~L_\pm\eqno(5)$$
$$L_+~L_-~-~L_-~L_+~=~[~2~L_0~]\eqno(6)$$
$$(L_0)^+~=~L_0~~,~~(L_\pm)^+~=~L_\mp\eqno(7)$$
where
$$[x]~=~{q^x~-~q^{-x}\over q-q^{-1}}~.\eqno(8)$$

The substitutions
$$L_0~=~J_0\eqno(9)$$
$$L_+~=~\left({p\over k}\right)^{{1\over2}~J_0~-~{1\over4}}~J_+\eqno(10)$$
$$L_-~=~J_-~\left({p\over k}\right)^{{1\over2}~J_0~-~{1\over4}}\eqno(11)$$
bring the defining relations (5-8) of the $SU_q~(2)$ algebra ( the
$L$ algebra ) into the relations (1-4) satisfied by the two
parameters $SU_{p,k}(2)$ algebra ( the $J$ algebra ). In other
words, the $SU_{p,k}(2)$ algebra is generated by the $SU_q(2)$
algebra. This fact leads to some relations between the
quantities appearing in the representations of the $SU_{p,k}(2)$
and the corresponding ones in the representations of the
$SU_q(2)$ algebra.

{\it First remark}: The Casimir operator $C_q$ of the $L$
algebra defined as
$$C_q~=~L_-~L_+~+~[L_0]~[L_0~+~1]\eqno(12)$$
is identical with the Casimir operator $C_{p,k}$ of the $J$ algebra
because the generators $J_{0,\pm}$ are in fact functions of $L_{0,\pm}$.
It results that $L_{0,\pm}$ as well as $J_{0,\pm}$ commute with
$C_q$ which means that $C_q$ is an invariant of the $J$ algebra
and hence
$$C_{p,k}~=~C_q~.\eqno(13)$$

{\it Second remark}: The eigenstates $\vert lm>$ of the $(2l+1)$
dimensional irreducible representation of the $L$ algebra are in
the same time eigenstates of the $J$ algebra as a consequence of
the fact that these states are eigenstates of the Casimir
operator $C_{p,k}~=~C_q$ and of the generator $L_0~=~ J_0$.

{\it Third remark}: The matrix elements of $J_\pm$ are equal with
those of $L_\pm$ up to a factor which can be easily derived
from the equations (9-11) and the second remark.
$$<l~m+1\vert J_+~\vert l~m>~=~\left({p\over k}\right)^{-{1\over2}m-
{1\over4}}~\left([l-m]~[l+m+1]\right)^{{1\over2}}\eqno(14)$$
$$<l~m-1\vert~J_-~\vert l~m>~=~\left({p\over k}\right)^{-{1\over2}m+
{1\over4}}~\left([l+m]~[l-m+1]\right)^{{1\over2}}~.\eqno(15)$$

{\it Fourth remark}: If $J_{0\pm}(1)$ is a set of generators
commuting with a second set of generators $J_{0\pm}(2)$, then one
can find a third set of generators $J_{0\pm}(1,2)$ defined as
follows
$$J_0(1,2)~=~J_0(1)~+~J_0(2)\eqno(16)$$
$$J_\pm(1,2)~=~J_\pm(1)~p^{-J_0(2)}~+~k^{J_0(1)}~L_\pm(2)\eqno(17)$$
which have the commutation relations and the hermitian
properties (1-4).

The corresponding generators of the $L$ algebra defined by
$$L_0(1,2)~=~L_0(1)~+~L_0(2)\eqno(18)$$
$$L_\pm(1,2)~=~L_\pm(1)~q^{-L_0(2)}~+~q^{L_0(1)}~L_\pm(2)\eqno(19)$$
can be transformed into $J_0(1,2)$ and $J_\pm(1,2)$ by using the
substitution (9-11).

{\it Fifth remark}: The Clebsch Gordan coefficients of the $L$
algebra are identical with the Clebsch Gordan coefficients of
the $J$ algebra because the eigenstates $\vert l_1~l_2~lm>$ are
labeled by the eigenvalues of the same set of operators in
both cases.

This remark remains valid for all recoupling coefficients.

{\it Sixth remark}: The $(2\lambda~+~1)$ components
$T_{\lambda\mu}$ of an irreducible tensor in the $L$ algebra
remain unmodified in the $J$ algebra. The relations
$$\left(~L_\pm~T_{\lambda\mu}~-~q^\mu~T_{\lambda\mu}~L_\pm~\right)
{}~q^{L_0}~=~\left([\lambda~\mp~\mu]~[\lambda~\pm~\mu~+1]\right)^{1\over2}
{}~T_{\lambda\mu\pm1}\eqno(20)$$
which define a tensor in the $L$ algebra [2] are changed into
$$\left(J_\pm~T_{\lambda\mu}~-~k^\mu~T_{\lambda\mu}~J_\pm~\right)~p^{J_0}~=~
\left({p\over k}\right)^{\lambda-\mu-{1\over2}\pm{1\over2}}~
\left([[\lambda\mp\mu]]~[[\lambda\pm\mu+1]]\right)^{1\over2}~
T_{\lambda\mu\pm1}\eqno(21)$$
for the $J$ algebra.

The fact that one can go from the equation (20) to equation (21)
by using the substitutions (9-11) shows that $T_{\lambda\mu}$ is
the same in both algebras.
\vskip 1.5truecm
{\bf References}

\item[1] Schirrmacher A, Wess J and Zumino B 1991 Z. Phys. C
{\bf49} 317

Ogievetsky O and Wess J 1991 Z. Phys. C {\bf50} 123

Smirnov Yu F and Wehrhahn R F 1992 J. Phys. A: Math. Gen. {\bf 25}
5563
\item[2] Biedenharn L C and Tarlini M 1990 Lett. Math. Phys.
{\bf20} 271
\bye